\documentclass[aps, prl,10pt,twocolumn,superscriptaddress,nofootinbib]{revtex4-1}
\usepackage{mathrsfs, amssymb, amsmath}  
\usepackage{epsfig, cancel}
\usepackage{latexsym}
\usepackage{natbib, comment}
\usepackage{url}
\usepackage{dcolumn}
\usepackage{multirow}
\usepackage{color}
\usepackage{cancel}
\usepackage{soul}
\usepackage[normalem]{ulem}
\usepackage{amsfonts,amssymb,amsmath, txfonts}
\usepackage{graphicx,epsfig}
\usepackage{psfrag}
\usepackage{hyperref}
\hypersetup{colorlinks=true}
\usepackage{mathtools}
\usepackage{enumitem}
\usepackage{float}
\usepackage[dvipsnames]{xcolor}
\usepackage{xcolor}
\hypersetup{ linktoc=all,
    colorlinks, linkcolor={blue},
    citecolor={darkred}, urlcolor={darkgreen}
}
\definecolor{rosy}{RGB}{230,235,252}
\definecolor{myframetitle}{RGB}{90,89,170}
\definecolor{myblocktitle}{RGB}{140,185,249}
\definecolor{mytitle}{RGB}{10,80,26}

\definecolor{darkgreen}{RGB}{27,130,45}
\definecolor{darkblue}{rgb}{0,0,0.3}
\definecolor{darkred}{rgb}{0.7,0,0}

\definecolor{light gray}{RGB}{220,220,220}
\definecolor{dark purple}{RGB}{108,0,217}
\definecolor{pink}{RGB}{190,20,100}
\definecolor{orang}{RGB}{193,63,0}
\definecolor{green}{RGB}{11,98,17}
\definecolor{darkpink}{RGB}{153,0,76}
\definecolor{bluegreen}{RGB}{0,102,102}
\definecolor{greenlagan}{RGB}{0,102,0}
\definecolor{redgreen}{RGB}{102,102,0}
\definecolor{Redgreen}{RGB}{153,76,0}
\definecolor{vividviolet}{rgb}{0.62, 0.0, 1.0}
\definecolor{amaranth}{rgb}{0.9, 0.17, 0.31}
\definecolor{palatinateblue}{rgb}{0.15, 0.23, 0.89}
\definecolor{brightpink}{rgb}{1.0, 0.0, 0.5}
\definecolor{cornflowerblue}{rgb}{0.39, 0.58, 0.93}
\definecolor{deepcarminepink}{rgb}{0.94, 0.19, 0.22}
\definecolor{radicalred}{rgb}{1.0, 0.21, 0.37}

\def\H0{{\text{H}\hspace*{-2.05mm}\text{H} 0\hspace*{-1.35mm}0\ }}

\def\be{\begin{equation}}
\def\ee{\end{equation}}
\def\beq{\begin{equation}}
\def\eeq{\end{equation}}
\def\bea{\begin{eqnarray}}
\def\eea{\end{eqnarray}}
\newcommand{\dd}{\textrm{d}}

\begin{document}

\title{DESI and SNe: Dynamical Dark Energy, $\Omega_m$ Tension or Systematics?}

\author{Eoin \'O Colg\'ain}
\affiliation{Atlantic Technological University, Ash Lane, Sligo, Ireland}
\author{M. M. Sheikh-Jabbari} 
\affiliation{School of Physics, Institute for Research in Fundamental Sciences (IPM), P.O.Box 19395-5531, Tehran, Iran}

\begin{abstract}
Dark Energy Spectroscopic Instrument (DESI) observations, when combined with Cosmic Microwave Background (CMB) and Type Ia supernovae (SNe), have led to statistically significant dynamical dark energy (DDE) claims. In this letter we reconstruct the $\Lambda$CDM parameter $\Omega_m$ from the $w_0 w_a$CDM cosmologies advocated by the DESI collaboration. We identify i) a mild increasing $\Omega_m$ trend at high redshift and ii) a sharp departure from $\Lambda$CDM at low redshift. The latter, which is statistically significant,  is driven by SNe that are $1.9 \sigma- 2.5 \sigma$ discrepant with DESI full-shape galaxy clustering in overlapping redshift ranges. We identify a low redshift subsample of the Dark Energy Survey (DES) SNe sample that is discrepant with DESI at $3.4 \sigma$ despite both observables probing the same effective redshift. SNe and BAO/full-shape modeling should not disagree on $\Omega_m$ at the same effective redshift. This ``$\Omega_m$ tension''  most likely points to unexplored systematics. In general, any \textit{bona fide} DDE signal should be confirmed independently across observables, so the DDE claims are premature.

\end{abstract}

\maketitle

\section{Introduction}
Confronted by persistent $\Lambda$CDM tensions \cite{DiValentino:2021izs, Perivolaropoulos:2021jda, Abdalla:2022yfr}, notably $H_0$ \cite{Planck:2018vyg, Riess:2021jrx, Freedman:2021ahq, Pesce:2020xfe, Blakeslee:2021rqi, Kourkchi:2020iyz} and $S_8$ tensions \cite{Planck:2018vyg, Heymans:2013fya, Joudaki:2016mvz, DES:2017qwj, HSC:2018mrq, KiDS:2020suj, DES:2021wwk}, the simplest deduction one makes is that the $\Lambda$CDM cosmological parameters cannot be constant. More precisely, if the tensions are physical in origin - not due to systematics -  the model fitting parameters must exhibit redshift-dependence; the model has broken down. Inspired by $H_0$ tension \cite{Planck:2018vyg, Riess:2021jrx, Freedman:2021ahq, Pesce:2020xfe, Blakeslee:2021rqi, Kourkchi:2020iyz} and strongly lensed quasars \cite{Wong:2019kwg, Millon:2019slk}, in 2020 it was argued that $H_0$ decreases with effective redshift \cite{Krishnan:2020obg}. Subsequently, a mathematical explanation was given \cite{Krishnan:2020vaf} and the observation was confirmed by independent groups, independent datasets and independent methods \cite{Dainotti:2021pqg, Dainotti:2022bzg, Hu:2022kes, Colgain:2022nlb, Colgain:2022rxy, Jia:2022ycc, jia2024uncorrelated}. 

In 2022, despite no discussion of an ``$\Omega_m$ tension" at the time, leveraging a well-recognized anti-correlation between the Hubble constant $H_0$ and the matter density parameter $\Omega_m$ in the $\Lambda$CDM model, coupled with an existing quasar anomaly \cite{Risaliti:2018reu, Lusso:2020pdb}, it was argued that  $\Omega_m$ increases as we increase the effective redshift \cite{Colgain:2022nlb, Colgain:2022rxy} (see also \cite{Pourojaghi:2022zrh, Pasten:2023rpc, Huang:2024erq, Colgain:2024clf, Huang:2024gfw}). Note, if $\Omega_m$ is not a constant, then the $S_8$ anomaly/tension, a discrepancy in $S_8 := \sigma_8 \sqrt{\Omega_m/0.3}$ deductions from different observations/datasets \cite{Planck:2018vyg, Heymans:2013fya, Joudaki:2016mvz, DES:2017qwj, HSC:2018mrq, KiDS:2020suj, DES:2021wwk}, is expected. In 2023, following precursor studies \cite{White:2021yvw, Garcia-Garcia:2021unp, Esposito:2022plo},\footnote{This is also predated by a late-time integrated Sachs-Wolfe effect anomaly \cite{DES:2018nlb, Kovacs:2021mnf}, which points to weaker/stronger growth at lower/higher redshifts.} it was proposed that $S_8$ increases with effective redshift \cite{Adil:2023jtu}. As noted in \cite{Akarsu:2024hsu}, this trend was subsequently observed in cluster number counts \cite{Artis:2024zag}, cross-correlations of DESI galaxies with CMB lensing \cite{ACT:2024nrz} and DESI full-shape galaxy clustering \cite{DESI:2024hhd, DESI:2024jis} (see also \cite{Tutusaus:2023aux, Manna:2024wak, Sailer:2024coh}).\footnote{Fig. 14 of \cite{DESI:2024jis} confirms that SDSS and DESI agree high redshift $f \sigma_8(z)$ constraints are larger than Planck. If low redshift constraints are lower (to be seen), DESI will confirm the claim in \cite{Adil:2023jtu}, which is based on archival data, using better quality data from a single survey/telescope. {See \cite{Mukherjee:2024pcg} for a recent striking realisation of the trends in SDSS data.}}  See \cite{Akarsu:2024qiq, Vagnozzi:2023nrq} for reviews of the redshift-dependent $\Lambda$CDM parameter claim.    

In 2024 DESI released baryon acoustic oscillation (BAO) results \cite{DESI:2024uvr, DESI:2024lzq, DESI:2024mwx}, which highlighted a discrepancy in the inferred $\Omega_m$  values from  BAO, CMB \cite{Planck:2018vyg} and {different combinations of} Type Ia SNe, including Pantheon+ \cite{Scolnic:2021amr, Brout:2022vxf}, Union3 \cite{ Rubin:2023ovl} and DES samples \cite{DES:2024jxu, DES:2024upw}. It has been claimed in \cite{DESI:2024mwx} that this ``$\Omega_m$ tension" {across} datasets can be modeled as a statistically significant dynamical dark energy (DDE)  parametrized through $w_0w_a$CDM model \cite{DESI:2024mwx}. Importantly, the signal persists when one replaces DESI BAO \cite{DESI:2024uvr, DESI:2024lzq, DESI:2024mwx} with DESI full-shape galaxy clustering \cite{DESI:2024hhd, DESI:2024jis}. Although one may eschew the $\Lambda$CDM model and focus on the $w_0w_a$CDM model or variants \cite{Giare:2024gpk, Wang:2024dka, Chan-GyungPark:2024mlx, Chan-GyungPark:2024brx, Chan-GyungPark:2025cri, Gao:2024ily}, field theories \cite{Tada:2024znt, Giare:2024smz, Bhattacharya:2024hep, Ramadan:2024kmn, zabat:2024wof, OShea:2024jjw, Heckman:2024apk, Li:2024qso, Ye:2024ywg, Wolf:2024stt, Bhattacharya:2024kxp, Ye:2024zpk, Li:2024qus}, cosmographic expansions \cite{Luongo:2024fww, Ghosh:2024kyd, Pourojaghi:2024bxa}, data reconstructions \cite{DESI:2024aqx, Mukherjee:2024ryz, Dinda:2024ktd, Jiang:2024xnu, Luongo:2024zhc} and wider physical implications \cite{Chakravarty:2024pec, Croker:2024jfg, Favale:2024sdq, Wang:2024hwd, Green:2024xbb, Giare:2024syw, Wang:2024sgo, Jiang:2024viw, Alfano:2024jqn, Lu:2024hvv, Arjona:2024dsr, Wang:2024tjd, Alfano:2024fzv, Pang:2024wul}, it is prudent to confirm that the DDE signal also manifests itself as a redshift-dependent $\Omega_m$ parameter in the $\Lambda$CDM model.\footnote{The DDE signal was also reported independently in non-DESI data \cite{Chan-GyungPark:2024mlx, Chan-GyungPark:2025cri} and the same logic applies. There is a prevailing assumption in the field that $\Omega_m$ is a constant \cite{Poulin:2024ken, Pedrotti:2024kpn}, even in the flat $\Lambda$CDM model, which would obviously be contradicted by a DDE signal.} To that end, we observed that a $\sim 2 \sigma$ DDE signal in the DES sample alone \cite{DES:2024jxu, DES:2024upw} can be translated into an increasing $\Omega_m$ trend with effective redshift \cite{Colgain:2024ksa}, otherwise one would arrive at a contradiction. 

This letter is inspired in part by a recent mapping \cite{Tang:2024lmo} of the \textit{mock data} based on the $w_0w_a$CDM cosmologies preferred by BAO+CMB+SNe \cite{DESI:2024mwx} back into the $\Lambda$CDM model. Unfortunately, the paper overlooks the evident redshift-dependence when \textit{observed data} is mapped back to the $\Lambda$CDM model, as outlined in the opening 2 paragraphs. This letter, in part, is intended to set the record straight. Concretely, we confirm that, irrespective of the SNe dataset, once the DDE $w_0w_a$CDM  cosmologies are mapped back to the $\Omega_m$ parameter, one finds an increasing trend of $\Omega_m$ at higher redshifts. \textit{This provides visual confirmation of earlier claims \cite{Colgain:2022nlb, Colgain:2022rxy} (see also \cite{Colgain:2024xqj})}. 

More notably, one also finds a statistically significant departure from the constant $\Omega_m$ $\Lambda$CDM behavior at low redshift. As we argue, this departure is driven by the SNe samples; {in the appendix we show that removing SNe data inflates the errors, thus decreasing the statistical significance of the trend.} Compared to recent DESI full-shape galaxy clustering of a low redshift Bright Galaxy Survey (BGS) \cite{DESI:2024jis}, the discrepancy with SNe in closely related redshifts we estimate at the $\sim 2 \sigma$ level irrespective of the SNe dataset. Given the higher effective redshift of the DES sample, cropping the sample, we find that viewed through the $\Lambda$CDM lens, DES and DESI are \textit{inconsistent at $\sim 3.4 \sigma$ at the same effective redshift}. We note that our observations agree with \cite{Bousis:2024rnb, Poulin:2024ken, Huang:2024qno} that there is some disagreement between DESI and SNe data, and this disagreement appears to extend to larger distances. Our letter highlights the dangers of combining datasets in a DDE framework \cite{DESI:2024mwx, DESI:2024hhd} without first checking that the observables are consistent where they probe the same effective redshift. See \cite{Colgain:2024xqj} for earlier related comments.

\section{Analysis}
We begin our analysis by explaining the role of CMB when combined with observables, such as SNe and BAO that directly probe the late Universe.
The CMB constrains the shift parameters $R, l_A$ \cite{WMAP:2008lyn}:  
\begin{equation}
\begin{split}
&R = \sqrt{\Omega_m H_0^2} \frac{D_{M}(z_*)}{c}, \quad l_{A} =  (1+z_*) \frac{\pi D_{M}(z_*)}{r_s (z_*)}, \\
&D_{M}(z_*) := c \int_0^{z_*} \frac{\dd z}{H(z)}, \quad r_{s}(z_*) := \int_{z_*}^{\infty} \frac{c_s \dd z}{H(z)},      
\end{split}\end{equation}
where $r_s (z_*)$ is the comoving sound horizon and $D_{M}(z_*)$ is the comoving angular diameter distance defined at the photon-decoupling surface redshift, $z_*\approx 1100$. Although both $R$ and $l_{A}$ depend on the cosmological parameters, only $D_M(z_*)$ is sensitive to the late Universe Hubble parameter $H(z)$.   

As noted in \cite{Akarsu:2022lhx}, since $D_{M}(z_*)$ is an integrated quantity through to $z_* $, one can easily improve the fit by introducing features in $H(z)$, for example wiggles, where there are no direct observational constraints. This means  that CMB cannot drive a DDE signal, but it can respond to $\Lambda$CDM deviations driven by BAO and SNe. To stress this point further, we note from Table 3 of \cite{DESI:2024mwx} that DESI BAO combined with both CMB and a BBN prior leads to $\sim 2 \sigma$ deviation from $\Lambda$CDM,\footnote{Concretely, from DESI+CMB one gets $w_0 = -0.45^{+0.34}_{-0.21}, w_a = -1.79^{+0.48}_{-1.0}$  and from DESI+BBN+$\theta_*$ $w_0 = -0.53^{+0.42}_{-0.22}, w_a < -1.08$.} thus confirming that CMB is not driving the signal {(see \cite{Huang:2024qno} for earlier related comments).}\footnote{Some component of the DDE signal may still be due to CMB \cite{Giare:2024ocw, Chan-GyungPark:2024brx, Chan-GyungPark:2025cri}.} This is in line with mathematical expectations \cite{Akarsu:2022lhx}.
Having eliminated CMB as the origin of the DDE signal, we turn our focus to BAO and SNe.   

\begin{table*}[htb]
\centering 
\begin{tabular}{|c|c|c|c|c|}
\hline
\rule{0pt}{3ex} \textbf{Data} & \textbf{$H_0$ [km/s/Mpc]} & \textbf{$\Omega_m$} & \textbf{$w_0$} & \textbf{$w_a$} \\
\hline\hline 
\rule{0pt}{3ex} BAO+PantheonPlus+CMB & $68.03 \pm 0.77$ & $0.3085 \pm 0.0068$ & $-0.827 \pm 0.063$ & $-0.75^{+0.29}_{-0.25}$\\
\hline
\rule{0pt}{3ex} Full-Shape+PantheonPlus+CMB & $68.34 \pm 0.67$ & $0.3061 \pm 0.0064$ & $-0.858 \pm 0.061$ & $-0.68^{+0.27}_{-0.23}$ \\
\hline
\rule{0pt}{3ex} BAO+DES+CMB & $67.24 \pm 0.66$ & $0.3160 \pm 0.0065$ & $-0.727 \pm 0.067$ & $-1.05^{+0.31}_{-0.27}$ \\
\hline
\rule{0pt}{3ex} Full-Shape+DES+CMB & $67.48 \pm 0.62$ & $0.3142 \pm 0.0063$ & $-0.761 \pm 0.065$ & $-0.96^{+0.30}_{-0.26}$ \\
\hline
\end{tabular}
\caption{$w_0 w_a$CDM cosmologies preferred by different datasets.}
\label{tab:CPL}
\end{table*}

\begin{figure}[htb]
   \centering
\includegraphics[width=80mm]{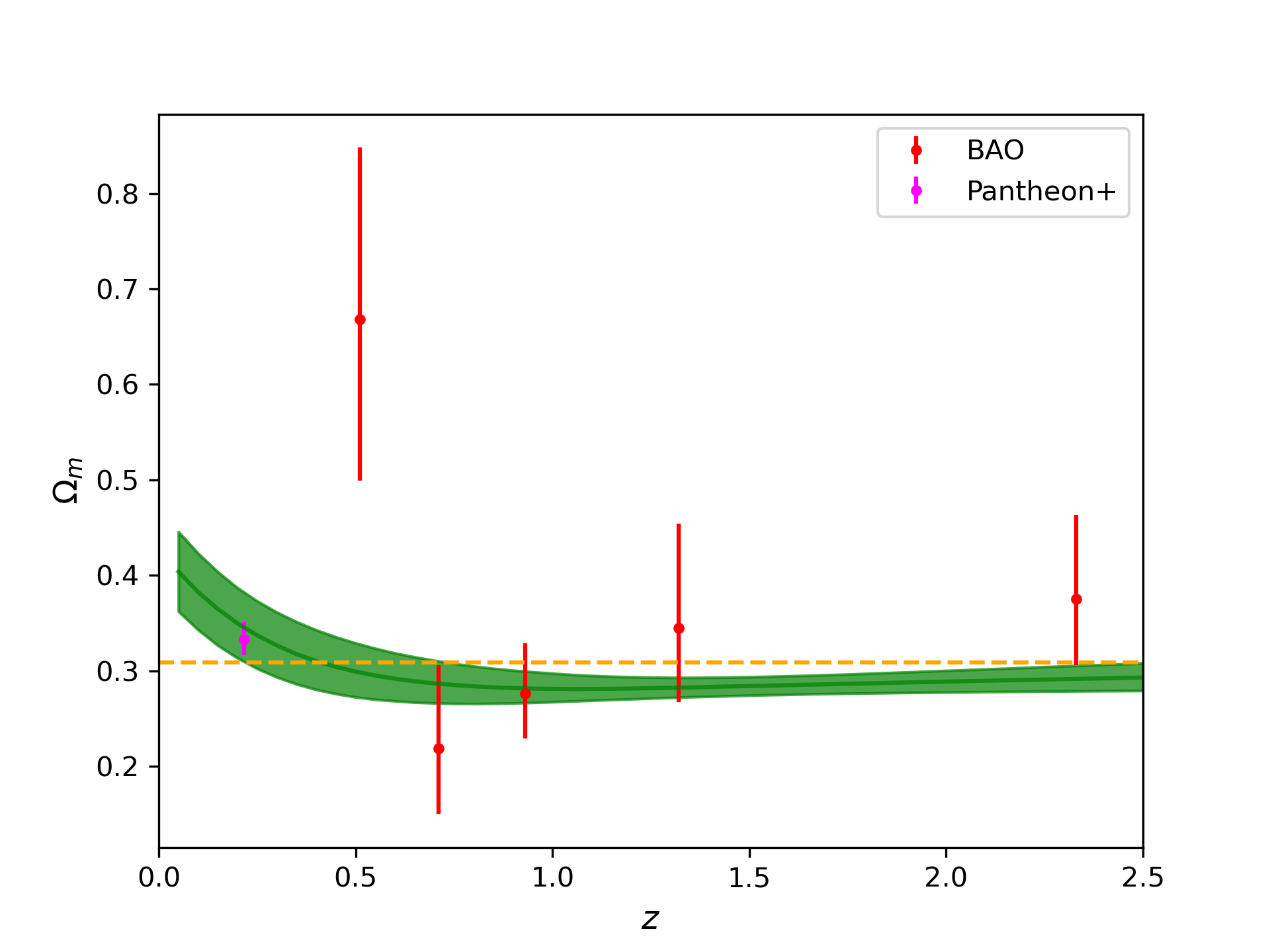}
\caption{Inferred values of $\Omega_m$ when the BAO+PantheonPlus+CMB $w_0w_a$CDM cosmology from Table \ref{tab:CPL} is mapped back to $\Lambda$CDM. Shaded green intervals correspond to $68 \%$ confidence intervals plotted alongside the Pantheon+ constraint and constraints inferred from DESI anisotropic BAO. The orange dashed line denotes the asymptotic value $\Omega_m = 0.3085$.}
\label{fig:Pantheon_bao} 
\end{figure}

\begin{figure}[htb]
   \centering
\includegraphics[width=80mm]{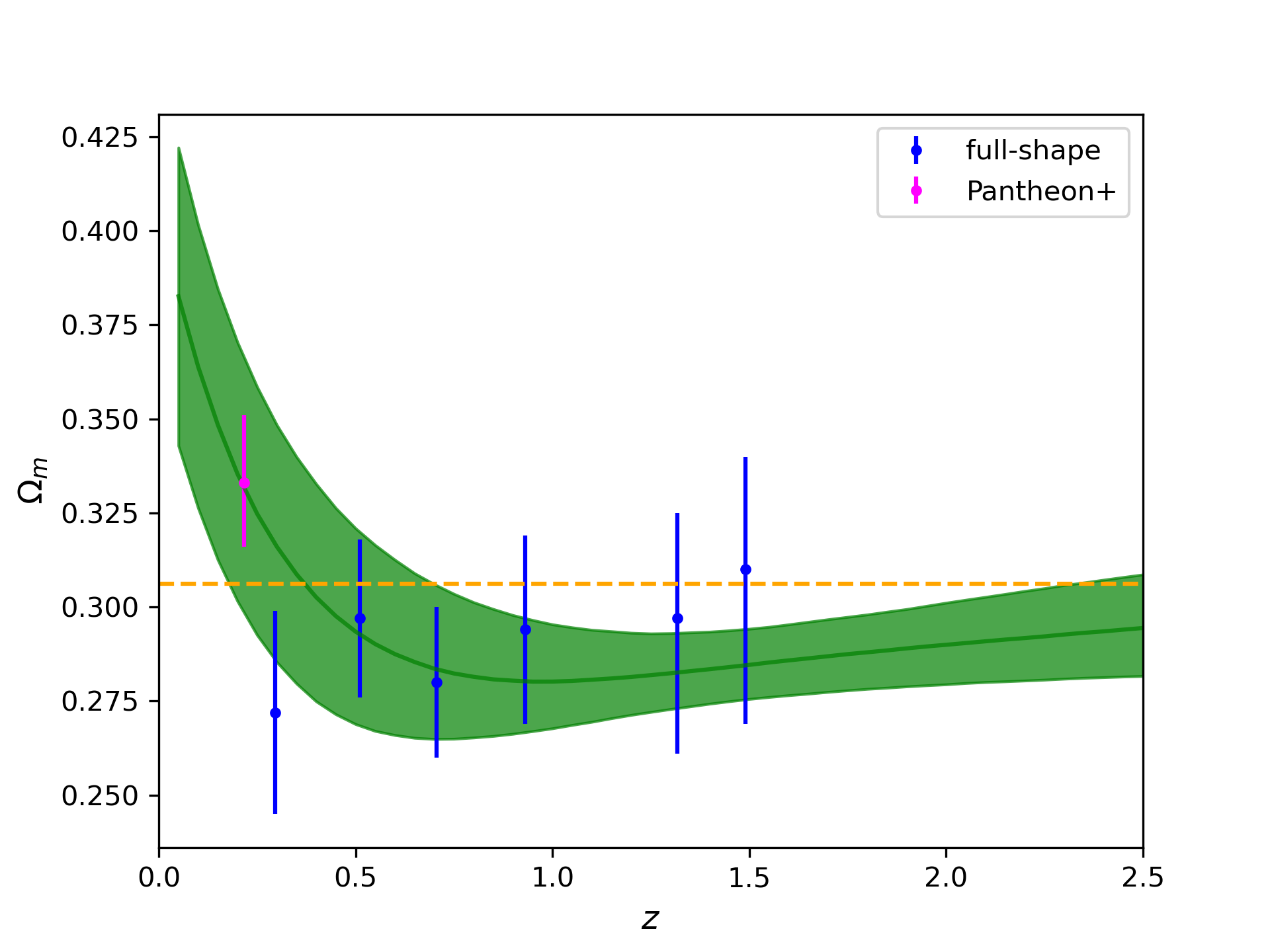}
\caption{Inferred values of $\Omega_m$ when the Full-Shape+PantheonPlus+CMB cosmology from Table \ref{tab:CPL} is mapped back to $\Lambda$CDM. Shaded green intervals correspond to $68 \%$ confidence intervals plotted alongside the Pantheon+ constraint and constraints inferred from DESI full-shape galaxy clustering. The orange dashed line denotes the asymptotic value $\Omega_m = 0.3061$.}
\label{fig:Pantheon_full} 
\end{figure}

\begin{figure}[htb]
   \centering
\includegraphics[width=80mm]{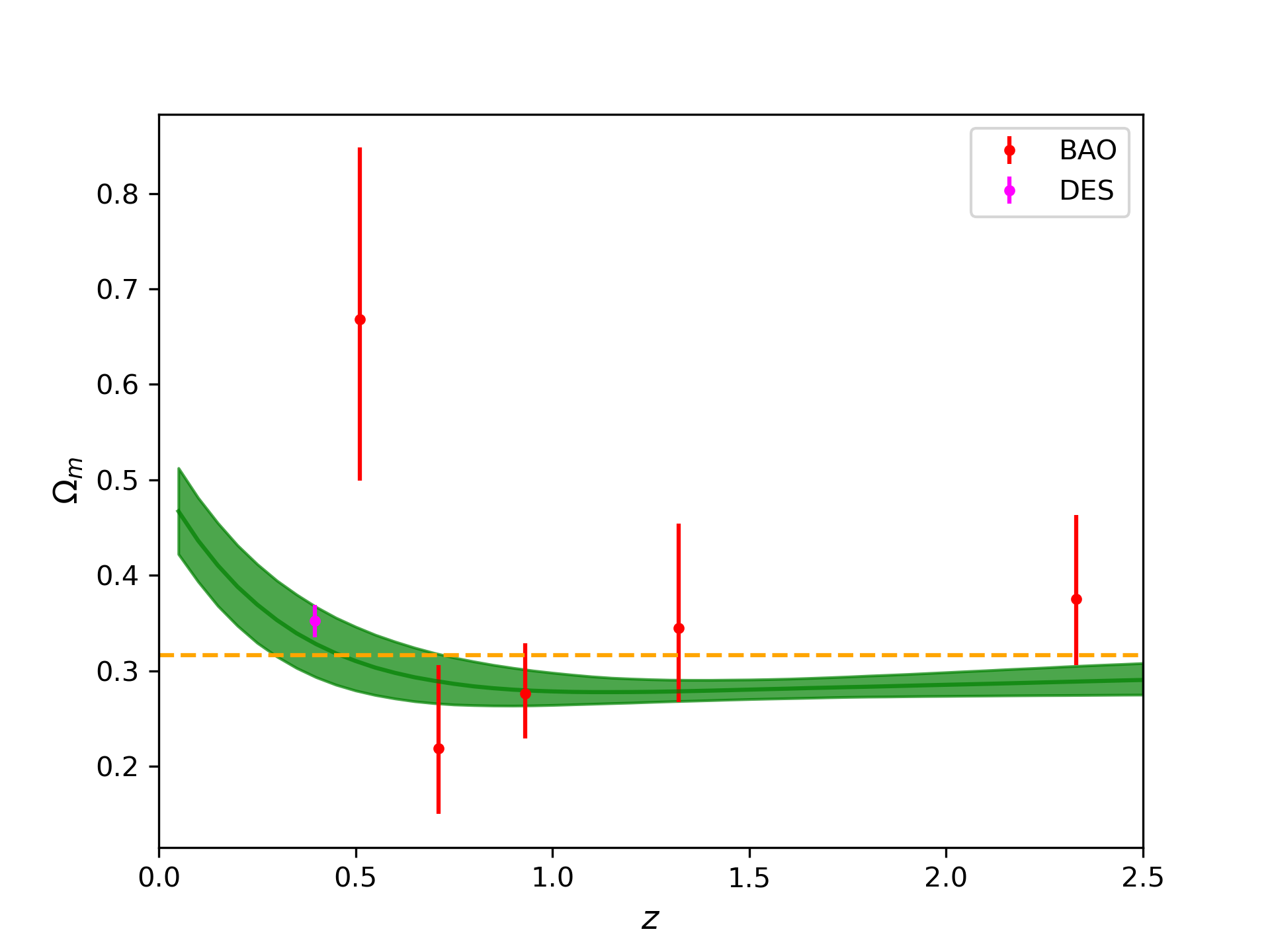}
\caption{Inferred values of $\Omega_m$ when the BAO+DES+CMB cosmology from Table \ref{tab:CPL} is mapped back to $\Lambda$CDM. Shaded green intervals correspond to $68 \%$ confidence intervals plotted alongside the DES constraint and constraints inferred from DESI anisotropic BAO. The orange dashed line denotes the asymptotic value $\Omega_m = 0.3160$. }
\label{fig:DES_bao} 
\end{figure}

\begin{figure}[htb]
   \centering
\includegraphics[width=80mm]{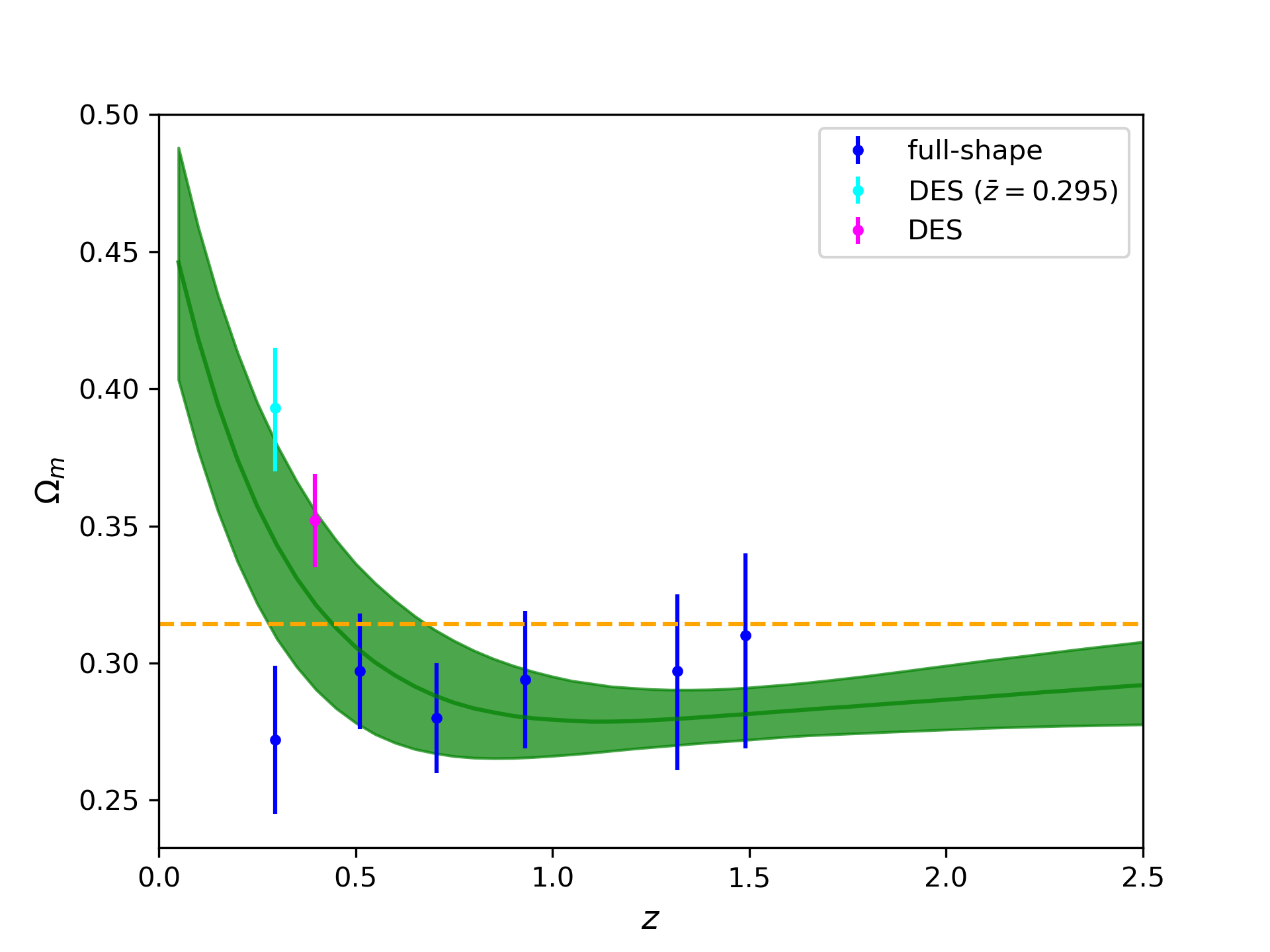}
\caption{Inferred values of $\Omega_m$ when the Full-Shape+DES+CMB cosmology from Table \ref{tab:CPL} is mapped back to $\Lambda$CDM. Shaded green intervals correspond to $68 \%$ confidence intervals plotted alongside the DES full sample constraint (magenta), DES subsample constraint (cyan) and constraints inferred from DESI full-fhape galaxy clustering. The orange dashed line denotes the asymptotic value $\Omega_m = 0.3142$.}
\label{fig:DES_full} 
\end{figure}

Given the best-fit $w_0w_a$CDM cosmologies in Table \ref{tab:CPL}, one can map the cosmologies back into the $\Lambda$CDM parameter $\Omega_m$ as follows.\footnote{While there may be other ways to do this, our methodology here is mathematically exact with no approximation.} First, we break up the redshift range $z \in [0.05, 2.5]$ into discrete values $z_i$ separated uniformly by $\Delta z = 0.05$. At each $z_i$ we generate 10,000 new values of $(H_0, \Omega_m, w_0, w_a)$ in normal distributions about the central values from Table \ref{tab:CPL} using the $68 \%$ confidence intervals (errors) as standard deviations. In the case of $w_a$, where the errors are not symmetric, for simplicity we symmetrize the errors by adopting the \textit{largest} error; this tweak can only make our analysis more conservative. From the resulting $(H_0, \Omega_m, w_0, w_a)$ array, one can construct a distribution of $D_{M}(z_i)$ and $D_{H}(z_i)$ values, where $D_{H}(z):= c/H(z)$. Next, following the methodology from \cite{Colgain:2022nlb, Colgain:2024xqj}, we translate these distributions into an $\Omega_m$ distribution by solving the equation 
\begin{equation}
\label{eq:ratio}
\begin{split}
        &\frac{D_{M}(z_i)}{D_H(z_i)} = E(z_i) \int_0^{z_i} \frac{\dd z}{E(z)} ,\\    &E(z) = \sqrt{1-\Omega_m + \Omega_m (1+z)^3}.
        \end{split}
\end{equation}
Note that $D_{M}(z_i)/{D_H(z_i)}$ ratio is only a function of $\Omega_m$ in the flat $\Lambda$CDM model, and more generally, does not depend on $H_0$. From the resulting $\Omega_m$ distribution we identify a median, $16^{\textrm{th}}$ and $84^{\textrm{th}}$ percentile at each $z_i$. This leads to the green curved bands or envelopes in Fig. \ref{fig:Pantheon_bao} - Fig. \ref{fig:DES_full}.  

At this point we have our first result. These observations are overlooked when one focuses on datasets without considering redshift, e. g. \cite{Tang:2024lmo}. Let us begin with the basics. Given that we have injected a $w_0w_a$CDM cosmology that is distinct from $\Lambda$CDM at the $\sim 2-4\sigma$ level, one should not expect a constant $\Omega_m$. In particular, the $\Omega_m$ curve shows a pronounced deviation from constant $\Omega_m$ at low redshifts with a less pronounced deviation from constant $\Omega_m$ at higher redshifts. Throughout, the orange dashed line denotes the $\Omega_m$ value from the relevant $w_0w_a$CDM  cosmology in Table \ref{tab:CPL} and we have confirmed in the appendix that the green curve converges from below to this value at higher redshifts, in line with expectations. The increasing $\Omega_m$ trend at higher redshifts is consistent with the claims from independent observables \cite{Colgain:2022nlb, Colgain:2022rxy}. The same trend is seen in the BAO in Fig. \ref{fig:Pantheon_bao} and Fig. \ref{fig:DES_bao}, once one ignores the obvious ``statistical fluctuation" outlier in luminous red galaxy (LRG) data at $z = 0.51$ \cite{Colgain:2024xqj} (see also DESI \cite{DESI:2024mwx} for related comments).\footnote{A prediction for DESI $\Lambda$CDM results that $\Omega_m$ increases with effective redshift was made in \cite{Colgain:2024xqj}. This is clearly borne out in i) the $\Omega_m$ reconstructed from the $w_0 w_a$CDM model and ii) DESI full-shape galaxy clustering results \cite{DESI:2024jis}, admittedly at low significance. On the latter point, the shift between the lowest $\Omega_m$ at the lowest redshift and the highest $\Omega_m$ at the highest redshift is only $\sim 0.8 \sigma$. Furthermore, it is clear that one can safely interpolate a horizontal line through the green curves for $z \gtrsim 0.4$, so any departure from $\Lambda$CDM behaviour {at higher redshifts} is currently marginal. Future DESI data may resolve this feature.} The reader should note that this statistical fluctuation has disappeared in DESI full-shape galaxy clustering results, as expected. The $z=0.51$ LRG outlier has been studied in subsequent papers \cite{Dinda:2024kjf, Wang:2024rjd, Wang:2024pui, Chudaykin:2024gol, Liu:2024gfy, Vilardi:2024cwq, Sapone:2024ltl}, where it has been stressed that the $z=0.71$ LRG constraint plays a key role driving the BAO+CMB+SNe DDE signal. This is presumably due to the fact that it represents the lowest $\Omega_m $ value in Fig. \ref{fig:Pantheon_bao} and Fig. \ref{fig:DES_bao}, thereby explaining the dip in $\Omega_m$ reconstructed from the $w_0 w_a$CDM model. {Our green curved bands help explain an observation in \cite{Tang:2024lmo} that CMB, BAO and SNe data mocked up on DESI $w_0 w_a$CDM cosmologies lead to a hierarchy of values $\Omega_m^{\textrm{SNe}} > \Omega_m^{\textrm{CMB}} > \Omega_m^{\textrm{BAO}}$, since the effective redshift of the BAO data is expected to coincide with the dip.}

We now focus on an important fact: the pronounced deviation from constant $\Omega_m$ at low redshift. {We note that the larger $\Omega_m > 0.3$ values are at odds with claims of smaller $\Omega_m < 0.3$ values at low redshift in the literature \cite{Sakr:2023hrl}.} It is obvious that this cannot be attributed to BAO, because we see the same feature in the full-shape modeling in Fig. \ref{fig:Pantheon_full} and Fig. \ref{fig:DES_full}. Moreover, even in the full-shape modeling, the green curved envelope fails to track the lowest redshift Bright Galaxy Sample (BGS) constraint \cite{DESI:2024jis}. Thus, one pinpoints SNe as the culprit {(see appendix where we remove SNe)}. In Fig. \ref{fig:Pantheon_bao} and Fig. \ref{fig:Pantheon_full} we plot the Pantheon+ constraint, $\Omega_m = 0.334 \pm 0.018$ \cite{Brout:2022vxf}, at the effective redshift, $\bar{z} = 0.217$, for the full SNe sample. Here, we define our effective redshift through a weighted sum of the errors: 
\begin{equation}
    \bar{z} = \frac{\sum_{i=1}^N z_i/ \sigma_i^2}{\sum_{i=1}^N 1/\sigma_i^2}. 
\end{equation}
Given that Pantheon+ has 741 from 1701 SNe with $z < 0.1$, it is expected that Pantheon+ has a low effective redshift. While not at the same effective redshift, the discrepancy between the Pantheon+ constraint in magenta in Fig. \ref{fig:Pantheon_full} and the BGS constraint at $\bar{z} = 0.295$, $\Omega_m = 0.272 \pm 0.027$ \cite{DESI:2024jis}, is $1.9 \sigma$. It is interesting to note that the Pantheon+ $\Omega_m$ value coincides with the median of the $\Omega_m$ values reconstructed from the corresponding $w_0w_a$CDM model. 

Replacing Pantheon+ SNe with DES allows us to take the comparison with full-shape modeling further. Although the DES sample overlaps with Pantheon+ through 194 low redshift SNe, $z \lesssim 0.1$, the DES full sample has a much higher effective redshift, $\bar{z} = 0.397$. We have plotted the full sample constraint, $\Omega_m = 0.352 \pm 0.017$ \cite{DES:2024jxu}, in magenta in Fig. \ref{fig:DES_bao} and Fig. \ref{fig:DES_full}, where it is clear that the sample explains the low redshift deviation from constant $\Omega_m$. The effective redshift is between $\bar{z} = 0.295$ and $\bar{z} = 0.510$ probed by BGS and LRG tracers \cite{DESI:2024jis}. The discrepancy between DES SNe and these observables are $\sim 2.5 \sigma$ and $\sim 2 \sigma$, respectively. {As an aside, it was noted in \cite{Gialamas:2024lyw} that removing the low redshift SNe from DES decreased the significance of the BAO+CMB+SNe DDE signal. From Table I of \cite{Colgain:2024ksa}, removing low redshift SNe reduces $\Omega_m \sim 0.35$ to $\Omega_m \sim 0.32$, thereby shifting the magenta constraint in Fig. \ref{fig:DES_bao} and Fig. \ref{fig:DES_full} closer to the orange dashed lines.}

\begin{figure}[htb]
   \centering
\includegraphics[width=80mm]{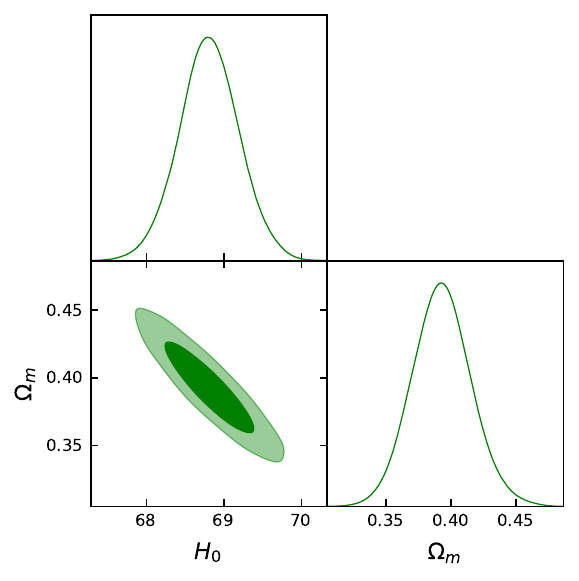}
\caption{Visual confirmation that DES SNe with $z < 0.597$ and an effective redshift $\bar{z} = 0.295$ lead to the constraint $\Omega_m = 0.393^{+0.022}_{-0.023}$.}
\label{fig:DESsubsample} 
\end{figure}

However, now we can remove the higher redshift DES SNe to construct a sample with the same effective redshift as the BGS $\Omega_m$ constraint. Concretely, we restrict the DES sample to $z < 0.597$ to get the same effective redshift $\bar{z} = 0.295$. The resulting DES constraint is $\Omega_m = 0.393^{+0.022}_{-0.023}$ from Fig. \ref{fig:DESsubsample}, which is $3.4 \sigma$ removed from BGS. \textit{We stress that the DES sample and DESI full-shape constraints are in tension where the samples share the same effective redshift.} This highlights the obvious shortcoming of glibly combining samples into a DDE framework, such as the $w_0w_a$CDM model \cite{DESI:2024mwx, DESI:2024hhd}. {As we explain in the appendix, one can translate this $\Omega_m$ tension back to distances using the $\Lambda$CDM cosmology, thereby confirming  DESI and DES disagreement on the distance at $\bar{z} = 0.295$. } 

{As a final comment, it is clear that full-shape modeling of galaxy clusters has led to $\Omega_m$ constraints that are more consistent with $\Lambda$CDM, modulo the residual increasing $\Omega_m$ trend with redshift. In principle, the conspicuous BAO outlier at $z = 0.51$ in Fig. \ref{fig:Pantheon_bao} and Fig. \ref{fig:DES_bao} can drive the DDE signal for BAO+CMB from Table 3 of \cite{DESI:2024mwx} (see \cite{Colgain:2024ksa}), {as explained in the appendix}. The analogous data combination for full-shape+CMB does not appear in Table 2 of \cite{DESI:2024hhd}, but if the results could be found, then this may add further support to our claim here that SNe drive the greatest deviation from constant $\Omega_m$ $\Lambda$CDM behavior. It would be most transparent to remove the SNe from Fig. \ref{fig:Pantheon_full} and Fig. \ref{fig:DES_full} to see if the low redshift deviation persists.}

\section{Discussion}
The DESI+CMB+SNe DDE signals raise perplexing theoretical questions. First and foremost, any dark energy model that crosses the phantom divide $w = -1$ is theoretically more of a challenge \cite{Vikman:2004dc}, which would consign quintessence \cite{Ratra:1987rm, Wetterich:1987fm, Copeland:2006wr}, the simplest field theory of dark energy, to the bin. Secondly, we note that a high local $H_0$ \cite{Riess:2021jrx, Freedman:2021ahq, Pesce:2020xfe, Blakeslee:2021rqi, Kourkchi:2020iyz} is also challenging to any dark energy field theory with an equation of state $w(z=0) = w_0 > -1$ \cite{Banerjee:2020xcn, Lee:2022cyh}. This statement places the DESI DDE claim at odds with $H_0$ tension. Even within the Horndeski class, this would push one to more exotic models in the class \cite{DeFelice:2011bh, Matsumoto:2017qil}. 

As advocated recently in \cite{Tang:2024lmo}, it is instructive to map the DDE $w_0w_a$CDM cosmologies preferred by DESI+CMB+SNe back into the $\Lambda$CDM model. While \cite{Tang:2024lmo} is a step in the right direction, it misses the point that the focus should be on effective redshift, rather than on different datasets. We know this because,  \textit{within the existing literature, mapping observed data} back into the parameters of the $\Lambda$CDM model has led to trends whereby $H_0$ decreases with effective redshift \cite{Krishnan:2020obg,Krishnan:2020vaf,Dainotti:2022bzg, Hu:2022kes, Colgain:2022nlb, Colgain:2022rxy, Jia:2022ycc, jia2024uncorrelated}, $\Omega_m$ increases with effective redshift \cite{Colgain:2022nlb, Colgain:2022rxy, Colgain:2024ksa, Colgain:2024xqj} (a trend evident in DESI full-shape modeling constraints at low significance) and $\sigma_8/S_8$ increases with effective redshift \cite{White:2021yvw, Garcia-Garcia:2021unp, Esposito:2022plo, Adil:2023jtu, Akarsu:2024hsu, Artis:2024zag, ACT:2024nrz, Tutusaus:2023aux, Manna:2024wak, Sailer:2024coh}. As initially argued for $H_0$ \cite{Krishnan:2020vaf}, these trends are expected signatures of $\Lambda$CDM model breakdown. {As a further comment, despite (\ref{eq:ratio}) being $H_0$ dependent, it is inevitably anti-correlated with $\Omega_m$ in the $\Lambda$CDM model. Since a DDE signal translates into a time-dependent $\Omega_m$ (the point of this letter), a time-dependent $H_0$ anti-correlated with $\Omega_m$ is expected from any DDE signal when mapped back to the $\Lambda$CDM model.}

We uncovered two interesting redshift-dependent features in our $\Lambda$CDM $\Omega_m$ reconstructions of the $w_0w_a$CDM model preferred by DESI+CMB+SNe data. The first is that there is an increasing trend of $\Omega_m$ with redshift at higher redshifts. This is in line with trends reported in the literature \cite{Colgain:2022nlb, Colgain:2022rxy, Colgain:2024clf, Colgain:2024ksa, Colgain:2024xqj}. The second and more noticeable feature is the departure from $\Lambda$CDM behavior at low redshifts $z \lesssim 0.5$. As we have argued, this is primarily driven by SNe samples that possess low effective redshifts and not by CMB or DESI data. Recent observations from DESI full-shape modeling of galaxy cluster are shifted lower than SNe by $\sim 2 \sigma$ at similar effective redshift. At $\sim 2 \sigma$, this may not be such a glaring problem, but as we have shown, one can tailor the DES SNe sample by removing higher redshift SNe to construct a sample with the same effective redshift as the BGS constraint. The constraints are inconsistent at $> 3 \sigma$ at the same redshift, indicating that DES SNe and DESI full-shape modeling cannot be tracing the same $\Lambda$CDM Universe. The key take-home is that this inconsistency is far from obvious when one combines DESI observations with CMB and SNe in the context of the $w_0w_a$CDM model.

The big picture here is that it has been argued that $H_0$ and $S_8$ tensions point to a crisis in cosmology. As argued initially in \cite{Krishnan:2020vaf}, differences in cosmological parameters at different effective redshifts can be reconciled if the $\Lambda$CDM model is broken down. Nevertheless, differences in cosmological parameters between observables at the same effective redshift are more serious; they raise the possibility that the observables are not tracking the same physical Universe \footnote{See \cite{Dhawan:2024gqy} for a discussion of SNe systematics in a related context.}. {Ultimately, if there is a genuine DDE signal, BAO and SNe need to independently recover the result.} 

\section*{Acknowledgements}
We thank \"Ozg\"ur Akarsu, Purba Mukherjee and Anjan Sen for related discussions. This article/publication is based upon work from COST Action CA21136 – “Addressing observational tensions in cosmology with systematics and fundamental physics (CosmoVerse)”, supported by COST (European Cooperation in Science and Technology). The work of MMShJ is in part supported by the INSF grant No 4026712.

\appendix 

\section{{Appendix: Further Technical Comments}}

Here we analyze the effect of removing SNe data from Fig. \ref{fig:Pantheon_bao} and Fig. \ref{fig:DES_bao} and show that it greatly inflates the errors, while leaving a residual non-constant $\Omega_m$/DDE signal. To this end we use DESI BAO  and CMB results for $w_0 w_a$CDM constraints \cite{DESI:2024mwx},
\begin{equation}
    (H_0, \Omega_m, w_0, w_a) = (64.7^{+2.2}_{-3.3}, 0.344^{+0.032}_{-0.027}, -0.45^{+0.34}_{-0.21}, -1.79^{+0.48}_{-1.0}).\nonumber
\end{equation}  
We repeat the steps from the main text with the only difference that instead of generating $w_0 w_a$CDM cosmologies that are normally distributed with symmetric errors, following the DESI $S_8$ analysis in \cite{Akarsu:2024hsu}, we model ($H_0, \Omega_m, w_0, w_a$) values larger and smaller than the central values as different multivariate normals. This ensures that we recover the correct central values and antisymmetric errors. Removing SNe data, the errors become more antisymmetric, and as the antisymmetric errors become more pronounced, it is prudent to change the methodology in the main text.

\begin{figure}[htb]
   \centering
\includegraphics[width=80mm]{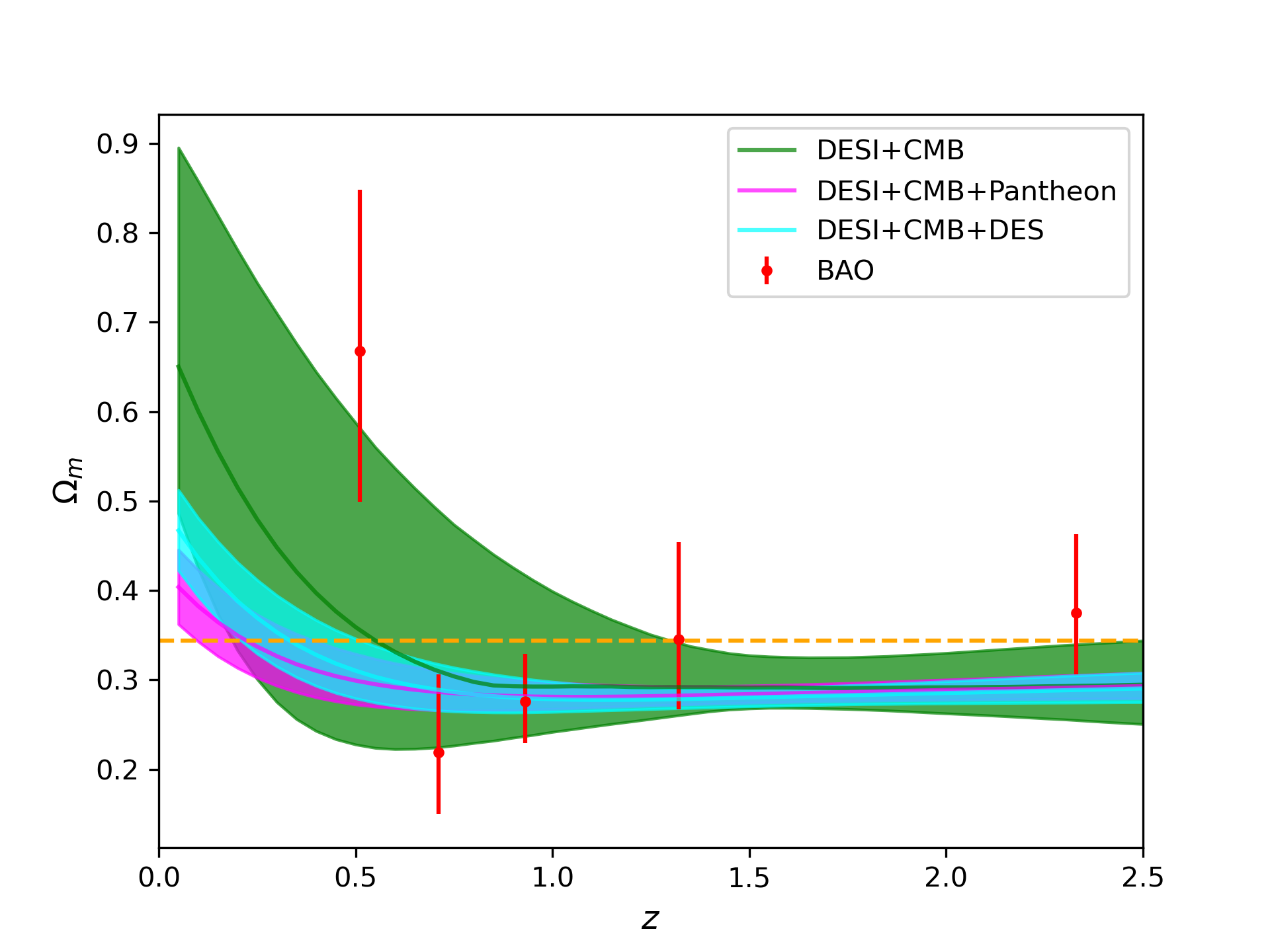}
\caption{{Confirmation that removing SNe data greatly inflates the 68\% confidence intervals. One can infer that $z=0.51$ LRG data drives any DDE signal in the absence of SNe.}}
\label{fig:cmb_bao} 
\end{figure}

\begin{figure}[htb]
   \centering
\includegraphics[width=80mm]{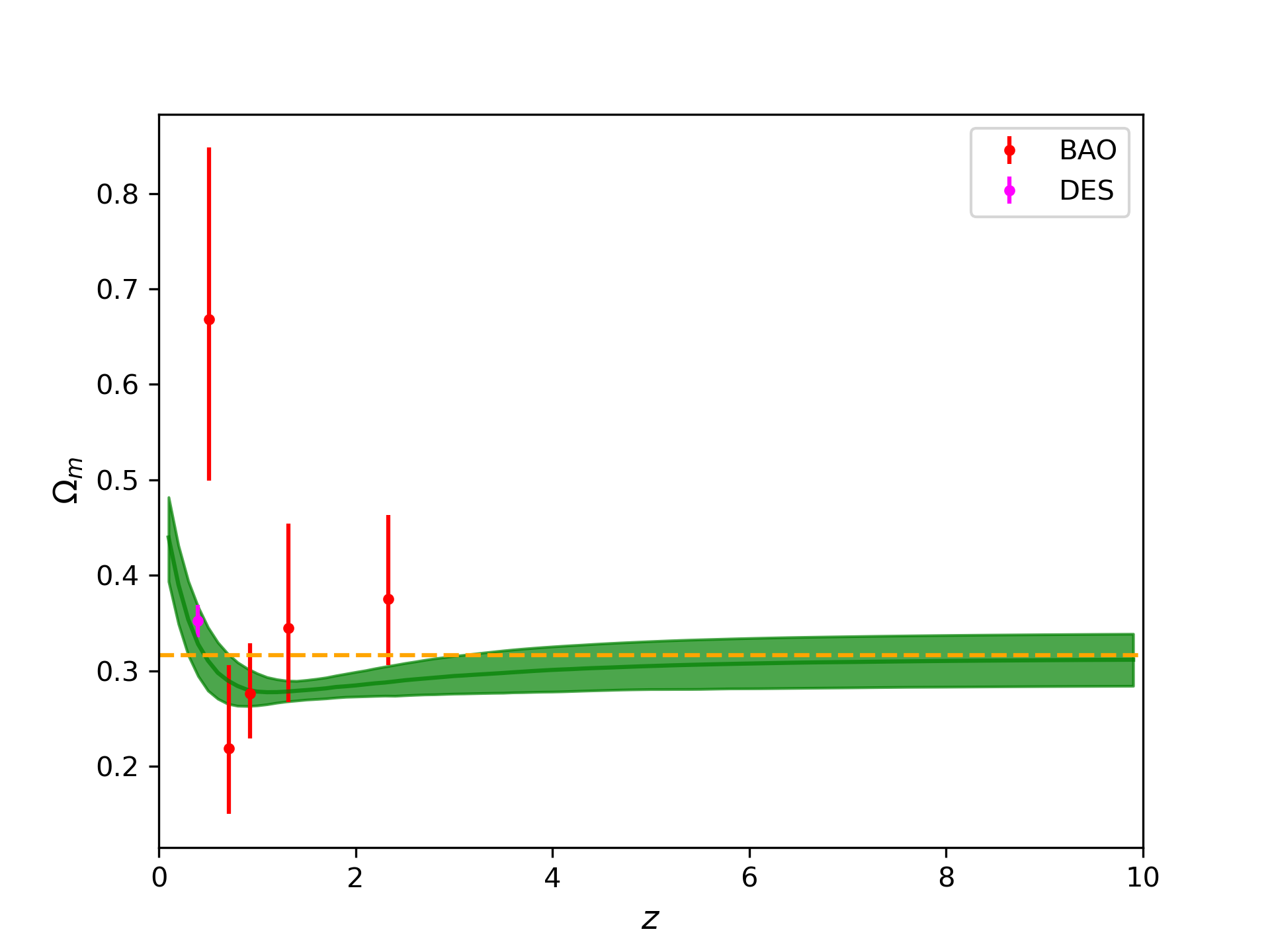}
\caption{{Same as Fig. \ref{fig:DES_bao}  but the green envelope extrapolated to $z=10$. The green curve converges from below to the input $w_0 w_a$CDM $\Omega_m$ value in dashed orange.}}
\label{fig:des_cmb_bao_extrapolate} 
\end{figure}

The $\Omega_m$ reconstruction of the DESI BAO+CMB $w_0 w_a$CDM model is shown in green in Fig. \ref{fig:cmb_bao} alongside the DESI BAO+CMB+SNe constraints. Once again, the dashed orange line denotes the asymptotic value $\Omega_m = 0.344$, where it is evident that $\Omega_m$ must increase with redshift to overcome a $\gtrsim 1 \sigma$ shift in $\Omega_m$ that is still present at $z = 2.5$. We confirm this with an explicit example in Fig. \ref{fig:des_cmb_bao_extrapolate}. We note that despite the errors inflating, there is still a pronounced deviation from $\Lambda$CDM behaviour (constant $\Omega_m$) at low redshift, which must in part be driven by LRG data at $z = 0.51$. This provides an alternative perspective on the results of  \cite{Colgain:2024ksa}, where in DESI data alone it was argued that LRG data at $z=0.51$ was responsible for the DDE signal. Strictly speaking, DESI BAO is combined with CMB, but as we have argued in the text, CMB constrains the integrated quantity $D_{M}(z_*)$, which still allows considerable freedom for $H(z)$ to deviate from Planck $\Lambda$CDM behaviour at specific redshifts. As a further comment, given that there is not an obvious outlier in DESI full-shape $\Omega_m$ constraints (see Fig. \ref{fig:Pantheon_full} and Fig. \ref{fig:DES_full}), if DESI provides the Full-Shape+CMB constraints in Table 2 of \cite{DESI:2024hhd},\footnote{The reader can compare with Table 3 of \cite{DESI:2024mwx} where DESI+CMB constraints for the $w_0 w_a$CDM model can be found.} it is likely we get no DDE signal or alternatively there is a constant $\Omega_m$ value that always inhabits the 68\% confidence envelope.

{Continuing the housekeeping, in Fig. \ref{fig:des_cmb_bao_extrapolate} we show that one can extrapolate the green $\Omega_m$ reconstruction from Fig. \ref{fig:DES_bao} to $z=10$ and confirm that the green curve converges from below to the $\Omega_m$ value of the input $w_0 w_a$CDM cosmology (orange dashed line). }

\begin{figure}[htb]
   \centering
\includegraphics[width=80mm]{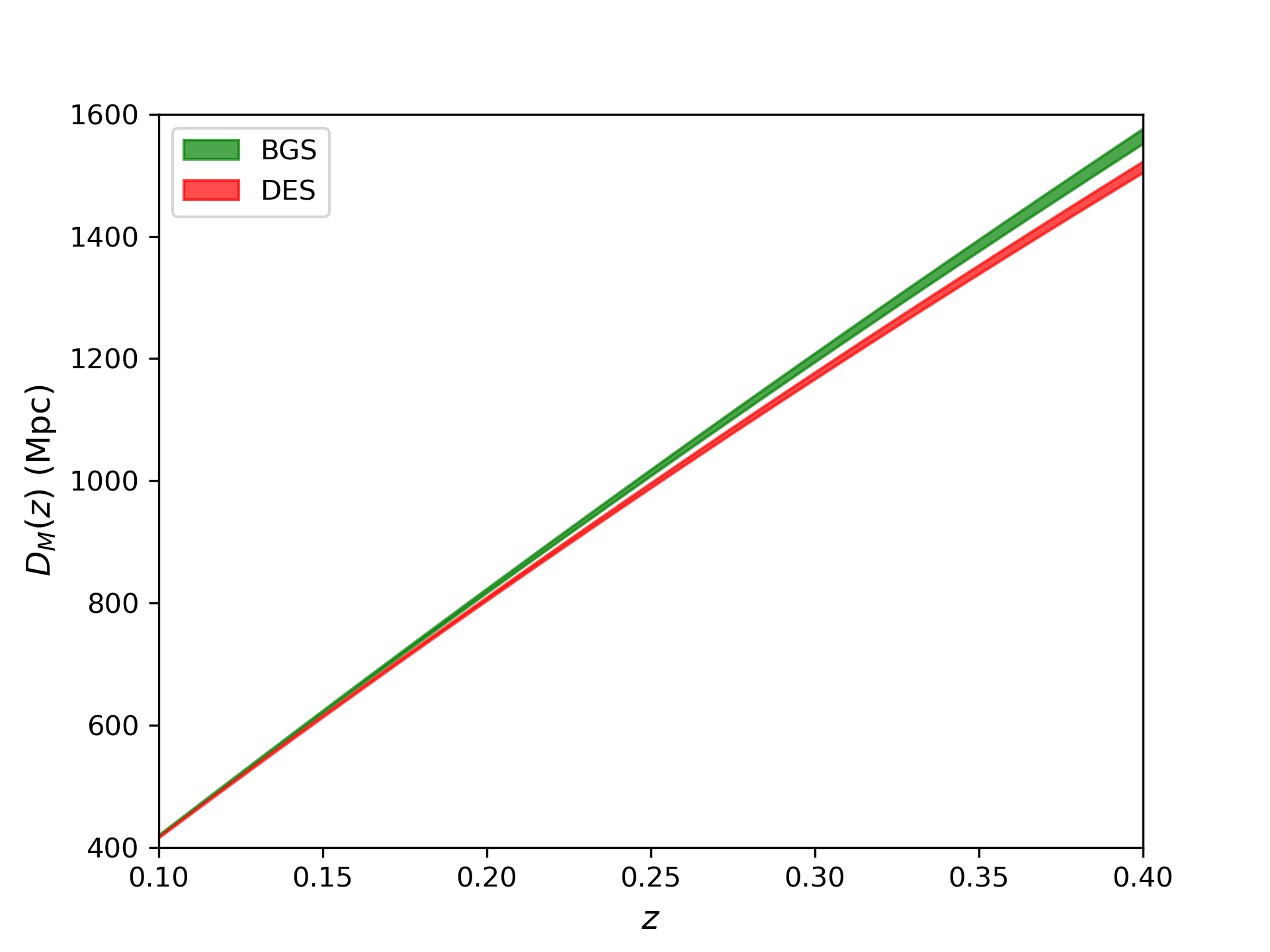}
\caption{Differences in $\Omega_m$ values between DES SNe and BGS at an effective redshift $\bar{z} = 0.295$ in Fig. \ref{fig:DES_full} translate into differences in the comoving distance in the overlapping redshift range $0. 1 \leq z \leq 0.4$.}
\label{fig:distances} 
\end{figure}

{Lastly, we spell out the implications for distances. Fig. \ref{fig:distances} confirms that the DESI BGS sample with full-shape galaxy clustering leads to different comoving distances $D_{M}(z)$ from DES SNe in the overlapping redshift range $0.1 \leq z \leq 0.4$. To construct these plots we have assumed $H_0 = 70$ km/s/Mpc and have generated $\Omega_m$ values in normal distributions, where we used $\Omega_m = 0.272 \pm 0.027$ for DESI BGS and $\Omega_m = 0.393 \pm 0.023$ for DES. The $68\%$ confidence intervals for the comoving distance is then constructed by isolating $16^{\textrm{th}}$ and $84^{\textrm{th}}$ percentiles for $D_{M}(z)$ at each $z$ sampled. As explained in the main text, both samples have the same effective redshift $\bar{z} = 0.295$. We note that DESI and DES cannot/should not disagree on distances in the same Universe. If systematics are not at play, this points to a problem with the $\Lambda$CDM model in overlapping redshift ranges. One can of course attempt to change the cosmological model to overcome the issue. We note that distance differences between DESI and SNe samples are highlighted independently in \cite{Bousis:2024rnb, Poulin:2024ken, Huang:2024qno} in binned analyses and no assumption is made on the cosmological model. Thus, changing the cosmological model is not expected to resolve the discrepancy.}

\bibliography{refs}

\end{document}